\documentstyle{article}
\def\beq{\begin{equation}}
\def\eeq{\end{equation}}
\def\nk{{\bf k}_{\nu}}
\def\mk{{\bf k}_{\mu}}
\begin{document}
\large
\title{WHAT IS THE VELOCITY OF GRAVITATIONAL WAVES?}
\author{Luciane R. de Freitas \\
Instituto de Fisica, UFRJ \\
and \\ M.Novello\\
Centro Brasileiro de Pesquisas Fisicas\\
Rio de Janeiro}
\date{ }
\maketitle

\begin{abstract}
We present a new field theory of gravity. It incorporates a great
part of General Relativity (GR) and can be interpreted in the standard
geometrical way like GR as far as the interaction of matter to
gravity is concerned. However, it differs from GR when treating gravity to
gravity interaction. The most crucial distinction concerns the
velocity of propagation of gravitational waves. Since there is a
large expectation that the detection of gravitational waves will
occur in the near future the question of which theory describes
Nature better will probably be settled soon.

\end{abstract}

\section{INTRODUCTION}
\protect\label{introduction}
\subsection{A. Introductory Remarks}

There is a general expectation concerning the possibility that we could
detect gravitational waves before the next century is born. Such
situation is based on the great number of new experimental devices that
many laboratories, throughout the world, are constructing. Many scientists
are going to become involved into this enterprise.
Thus the time is ripe to theoretical re-examination of gravity theory in
order to make predictions on the propagation of gravitational disturbances
that could be tested in the near future.

Although the great majority of physicists expect that the observation
of these waves will confirm the General Relativity prediction\footnote{The
recent spectacular success of the description of pulsar behavior, through
loss of gravitational energy, by General Relativity, increased enormously
the status of this theory.}, it seems worthwhile to remember that such an
expectation may just be one of those prejudgements that sometimes spread
into the scientific community. On the other hand, we would like to emphasize
that if this General Relativity  result is proven to be false, it does not
destroy the remaining and by far the major part of Einstein\rq s theory
that has already been experimentally proved.

The Equivalence Principle - which states that all kinds of matter (including
massless particles as photons) interact in an unique and same way with the
gravitational field - gave to Einstein the possibility to treat gravitational
phenomena as a sort of modification of the spacetime geometry.

Does this universal behavior occur also in the case of gravity-gravity
interaction? General Relativity makes an implicit hypothesis by means of which
the answer to this question is affirmative.
Although there is not even a single observational evidence that
this is true, the fact that gravity must carry energy as any other
field gave a strong motivation to believe that such an extrapolation
concerning gravity-gravity interaction should be valid.

This assumption, that still today remains beyond any real observation,
led to the most impressive result of General Relativity, that is, that
gravitational processes are nothing but a real universal modification of the
spacetime geometry.

However, if we want to keep observation as the true guide of
our analysis of Nature,
the whole actual situation can be summarized in two statements:
\begin{itemize}
 \item{As far as matter-to-gravity interaction is concerned, the
GR scheme of geometrization of gravity seems to be a very good procedure.}
 \item{There is not any single direct observational evidence that supports that
the self-interaction of gravity can be described in the same way.}
\end{itemize}

Thus, if we limit ourselves to the traditional scientific method of
submission of theory to observation, we must say that the universal
modification of the geometry proposed by GR is indeed an extrapolation
that is not still confirmed by experimental means, as far as the behavior
of gravity-gravity coupling is concerned.

We are thus led to take seriously into account the two excludent alternatives
concerning such gravity-to-gravity interaction, to wit:

\begin{itemize}
 \item{Gravity couples to gravity as any other form of energy.}
 \item{Gravity couples to gravity in a special way distinct from all
different forms of energy.}
\end{itemize}

In the first case, the universal modification of the geometry, as proposed
in General Relativity, becomes a natural scenario. Neverthless, the
complete absence of any experimental evidence that could help us
to solve this question tells us that the decision must be dictated by other
means, e.g., either by theoretical arguments\footnote{At most times
such a decision comes from theoretical prejudices.} or by some sort of
additional requirements imposed on the interaction mechanism.
Einstein\rq s hypothesis that gravitational interaction is the same not only
when gravity-matter interaction is concerned but also for gravity-gravity
processes, appeared to be the most natural way. In those early times,
the complete absence of observations postponed the decision
on this for the future. Later, in the fifties, the discovery of a simple way
to treat gravity in terms of a standard field theory \cite{Feynmann},
reducing the gap between GR and all other
field theories, made Einstein\rq s hypothesis less questionable.

The purpose of the present work is to show a new way to generate a field
theoretical description of gravity.

The fundamental property, that rests on the basis of our theory and
that distinguishes the present program of investigation from GR, can be
synthesized by the assertion that we will examine the possibility of
conciliating the universality of matter to gravity interaction but
considering that gravity to gravity coupling is somehow distinct.
Since, as we shall see, the most dramatic consequence of the theory we
propose here deals with the modification of the velocity of the
gravitational waves, we limit our analysis here to the exam of this question.

Just in order to support this statement we note that it is
a theoretical prejudice to argue that GR has already settled
this question: only observation could do this.

The new net result (concerning the propagation of gravitational waves) of our
model can be summarized by noting that in this theory gravitational waves
propagate in an effective geometry that is not the same as viewed by matter.
One can then ask "what is the true geometry of spacetime"? The answer will
be: it depends on the instrumentation we use to observe it.
All forms of energy, except the gravitational, measure the same universal
modification of curved spacetime. However, gravitational waves
behave as if they were imbedded in a distinct geometry.
What should the origin of this fact be? We postpone this question for latter
analysis.

Just to provide a simple example of a theory that displays these ideas
we will take as a model a non-linear theory proposed many years ago by
Born \cite{Born}
for spin-one field. We will apply a similar model for the spin-two case.
Although in the original
proposal of spin-one field such a non-linear theory appeared just as an
exotic one, in the case of spin-two that we treat here the
non-linearity is mandatory. We shall see in a subsequent section that this
situation is just a consequence of the fact that spin-two couples to the
energy-momentum tensor and the spin-one field couples to a conserved
current. The fact that gravity itself must, for consistency, have energy
makes the non-linearity a necessary requirement of any description of
gravity.

In the standard geometrical way such self-interaction is described by a
non-polinomial Lagrangian. In our re-exam of gravity from the point of view
of field theory we will take (just as an example of constructing a
coherent field theoretical model) the Born non-polinomial Electrodynamics
as a paradigm for the gravitational interaction.

Finally, let us note that although we concentrate here all our presentation
of the new theory only to the gravitational waves, we will make a few comments
in the final section, anticipating
some results that concern the properties of the new theory related to the
behavior of the gravitational field in two important situations, to wit:

\begin{itemize}
 \item{The static spherically symmetric field;}
 \item{Cosmology.}
\end{itemize}

We leave for a subsequent paper the proof that the standard tests of
gravitational processes are satisfied by our present theory.

\subsection{B. Synopsis}
The presentation of the paper is the following. In Section (\ref{def}) we
present the main definitions and symbols that we use.

In Section (\ref{geo})
we discuss briefly how and why Feynman \cite{Feynmann}, Deser
\cite{Deser} and others (see \cite{Gris} for a recent review) were led
to describe General Relativity in terms of a field theory only based on
the universality of gravitational interaction; we revise the Fierz linear
theory and propose a class of non-linear theory of spin-two field. We also
study the behaviour of the gravity energy-momentum tensor in the new theory.
We present a new derivation of Fierz equation of motion that is
worth of generalization in the non-linear case.

In the Section (\ref{model}) we propose a specific theory and, by analyzing
the evolution of the disturbances in this case, we show that
the gravitational waves propagate on the null cone of an effective
geometry distinct from that one seen by matter.

In Section (\ref{matter}) we
present the gravity-matter interaction process and following the
standard procedure we show how such an interaction can be described
in terms of a modification of the geometry of the spacetime, in the
same manner as it occurs in General Relativity.

Finally, in the last Section (\ref{conclusion}) we make some
comments and state future perspectives of our scenario on gravity.

\section{DEFINITIONS AND NOTATIONS}
\protect\label{def}

The auxiliary metric $\gamma_{\mu\nu}$ of Minkowski
geometry\footnote{We shall see in next sections that this metric is not
observable neither by matter nor by gravitational field.} is written
in an arbitrary system of coordinates in order to exhibit the general
covariance of the theory. We define the corresponding covariant derivative by

\beq
V_{\mu\hspace{0.1cm};\hspace{0.1cm}\nu} =
V_{\mu\hspace{0.1cm},\hspace{0.1cm}\nu} -
\Delta^{\alpha}_{\mu\nu} V_{\alpha}
\label{d0}
\eeq

in which
\beq
\Delta^{\alpha}_{\mu\nu} =
\frac{1}2 \gamma^{\alpha\beta}\hspace{0.1cm}(
\gamma_{\beta\mu\hspace{0.1cm},\nu} +
 \gamma_{\beta\nu\hspace{0.1cm},\mu} - \gamma_{\mu\nu\hspace{0.1cm},\beta}).
\label{d01}
\eeq

The associated curvature tensor vanishes identically that is

\beq
R_{\alpha\beta\mu\nu}(\gamma_{\epsilon\lambda}) = 0.
\label{d02}
\eeq

We define a three-index tensor $F_{\alpha\beta\mu}$, which we will call
the {\bf gravitational field}, in terms of the symmetric standard variable
$\varphi_{\mu\nu}$ (which will be treated as the potential) to
describe spin-two fields, by the expression

\beq
F_{\alpha\beta\mu} = \varphi_{\mu[\alpha;\beta]} +
\varphi_{,[\alpha}\gamma_{\beta]\mu} +
\gamma_{\mu[\alpha}{{\varphi_{\beta]}}^{\lambda}}_{;\lambda}.
\label{d1}
\eeq
where we are using the anti-symmetrization symbol $[\hspace{0.9mm}]$ like

\beq
[A, B] \equiv AB - BA.
\label{d4}
\eeq
We use an analogous form to the symmetrization symbol $(\hspace{0.9mm})$

\beq
(A, B) \equiv AB + BA,
\label{d5}
\eeq

{}From the above definition it follows that this quantity $F_{\alpha\beta\mu}$
is anti-symmetric in the first pair of
indices and obeys the cyclic identity, that is

\beq
F_{\alpha\mu\nu} + F_{\mu\alpha\nu} = 0.
\label{d2}
\eeq

\beq
F_{\alpha\mu\nu} + F_{\mu\nu\alpha} + F_{\nu\alpha\mu} = 0.
\label{d3}
\eeq

The trace of the tensor $F_{\alpha\beta\mu}$ is given by

\beq
F_{\mu} \equiv F_{\mu\alpha\beta} \gamma^{\alpha\beta} =
4({{\varphi_{\mu}}^{\lambda}}_{;\lambda} - \varphi_{\lambda}).
\label{d6}
\eeq

This allow us to re-write the expression of the field in the
form

\beq
F_{\alpha\beta\mu} = \varphi_{\mu[\alpha;\beta]} +
 \frac{1}{4} F_{[\alpha}
\gamma_{\beta]\mu}.
\label{d231}
\eeq

The equation of motion of the gravitational field will appear in
a more convenient form when written in terms of
an associated quantity $M_{\alpha\beta\mu}$ that has the same
symmetries as  $F_{\alpha\beta\mu}$ and is defined by

\beq
M_{\alpha\beta\mu} \equiv F_{\alpha\beta\mu} -
\frac{1}2 F_{\alpha} \gamma_{\beta\mu} + \frac{1}2
F_{\beta} \gamma_{\alpha\mu}.
\label{d7}
\eeq

See the Appendix for other properties of these quantities.

The quantity $\kappa$ represents Einstein\rq s constant, written in
terms of Newton\rq s constant $G_{N}$ and the velocity of light $c$
by the definition $$ \kappa = \frac{8\pi}{G_{N} c^{4}}. $$ We set $c
= 1.$

\section{FROM THE UNIVERSAL COUPLING OF MATTER TO GRAVITY TO THE
EINSTEIN GEOMETRIZATION SCHEME}
\label{geo}

In this section we will briefly review some points that were
used in order to implement the hypothesis of universality of gravity
interaction by the modification of the metrical properties of the
spacetime. We will be interested here not in its historical birth of
Einstein\rq s formulation but, instead in its equivalent
field theoretical formulation. In other
words, we will follow here the path that led from the linear field theory
of gravity to its non-linear processes and consequently to the
corresponding geometrization scheme\cite{Feynmann}, \cite{Deser}.

There is no better and simpler manner to describe this than the one
set by Feynmann in his lecture notes of 1962. Let us summarize such
standard procedure that led from the field theoretical to the
geometrical description of gravitational interaction.

The starting point is the linear massless spin-two field theory
(Fierz equation), that reads:
\beq
G_{\mu\nu}^{L} = -kT_{\mu\nu}
\protect\label{1}
\eeq
in which
\beq
G^{(L)}_{\mu\nu}\equiv \Box\phi_{\mu\nu} -\phi_{\ \mu\mid\alpha\nu}^{\alpha}
 -\phi_{\ \nu\mid\alpha\mu}^{\alpha} +\phi_{\ \alpha\mid\mu\nu}^{\alpha}
-\gamma_{\mu\nu}(\Box\phi_{\ \alpha}^{\alpha}
-\phi^{\alpha\beta}_{\ \ \mid\alpha\beta}).
\protect\label{gminilinear}
\eeq
The quantity $G_{\mu\nu}^{L}$ is divergence-free. This implies
that, for compatibility, one must impose the condition
that the energy-momentum tensor $T_{\mu\nu}$ of matter should also be
divergenceless. Now, since the gravitational field contributes to the balance
of the conservation law through its own energy, this imposition faces a
difficulty since the matter energy-momentum tensor cannot be
separately conserved.

It is precisely at this point that the hypothesis that gravity-gravity
process follows the same type of behavior as matter-gravity interaction
acts as a guide to the choice of the gravitational contribution to the
source of $G_{\mu\nu}^{L}$. This means to add to the energy-momentum tensor
of matter, the corresponding energy-momentum tensor for the gravitational
field at the right-hand-side of equation (\ref{1}).

The idea is to proceed step by step. We start by adding to the
right-hand-side of equation(\ref{gminilinear}) the tensor
${T^{1}}_{\mu\nu}$,
that is the energy momentum tensor of the linear equation for gravity. As a
consequence we must add to the original Lagrangian an additional
term of higher order that yields, after variation, the term
${T^{1}}_{\mu\nu}$ to be added to the equation of motion. This generates
a new compatibility condition, which is solved by adding to the
right-hand-side of the equation of motion a new term of higher
order, ${T^{2}}_{\mu\nu}$. This will impose, once more, that a new term must
be added to the Lagrangian.
This process continues indefinitely, since at each step a new term must be
added
to the Lagrangian in order to achieve the compatibility lost at the
precedent lower order. To accomplish the task and to solve completely
the compatibility condition, we
must deal with a recurrence procedure such that yields an infinite series to
appear \cite{Feynmann}. It is precisely the summation of such infinite
series that can be described by the equivalent geometrical
formulation of General Relativity. Our purpose in the present work is to
re-analyse this field theoretical description and to show that the above
traditional procedure of searching compatibility is not unique.

\subsection{Fierz Linear Theory Revisited}
\label{fierz}

We will now show that the linear theory of spin-2 field can be
described using the invariants constructed with the gravitational
field $F_{\alpha\mu\nu}$. We will perform this simple exercise here just in
order to present the motivation for our further non-linear theory.

There are only two invariants that can be constructed with the
field.\footnote{We could use instead of invariant $A$ the one
construted with $C_{\alpha\mu\nu}$, the traceless part of
$F_{\alpha\mu\nu}$ which
employs its irreducible parts. For simplicity of comparison
to the traditional Fierz theory we decided here to make the above choice.}
They are:

$$ A \equiv F_{\alpha\mu\nu}\hspace{0.5mm} F^{\alpha\mu\nu} $$

$$ B  \equiv F_{\mu}\hspace{0.5mm} F^{\mu}. $$

General covariance imposes that the Lagrangian that one can construct to
describe the evolution of the gravitational field must be a
functional of these invariants, that is

$$ L = L (A, B).  $$
The linear theory for the spin-2 field is given by the action

\beq
S^{L} = \frac{1}{2} \int  \sqrt{-\gamma} \hspace{0.8mm}
( - A + \frac{3}{4} B) d^{4}x.
\protect\label{s1}
\eeq
in which $\gamma$ represents the determinant of $\gamma_{\mu\nu}$.

The proof of this assertion can be made either by a direct inspection
on the equation of motion obtained from this Lagrangian or by noting
that up to a total divergence we can write

\beq
S^{L} =  \int  \sqrt{-\gamma} \varphi^{\mu\nu} G_{\mu\nu}^{L} d^{4}x
\label{s2}
\eeq

Indeed, we have from eq. (\ref{s2}), up to a total divergence
\beq
S^{L} = - \frac{1}{2} \int  \sqrt{-\gamma} \varphi^{\mu\nu}
{M^{\lambda}}_{(\mu\nu);\lambda} =  \frac{1}{2} \int
\sqrt{-\gamma} \varphi^{\mu\nu;\lambda} M_{\lambda(\mu\nu)}
\eeq

As a consequence of a direct manipulation of this expression
we can show directly that indeed
\beq
\varphi^{\mu\nu;\lambda}  M_{\lambda(\mu\nu)} = - A + \frac{3}{4} B.
\eeq
This demonstrates our assertion. What we have learned from this simple
manipulation is that any theory that  provides Fierz linear equation
of motion in the weak field limit should reduce to the above combination of
the invariants $A$ and $B$. It is tempting then to examine those
theories that are functionals {\bf only} of this combination. We will limit
thus all our analysis only to this set of theories. Besides, in the
present paper we will consider a specific example of dynamics
represented by a Lagrangian that is constructed as a non-polinomial
functional of the field variables.

Before going into the exam of such a non-linear theory for the
gravitational field, let us make a very short r\'esum\'e of a typical
example of a class of non-linear spin-one theory. We shall see that many
properties of this example will have a deep analogy to the spin-two case.
This will be useful since it will act as a guide for the analysis of the more
complex case of the gravitational field.

\subsection{Non-Linear Spin-One Theory}
\label{non}

The dynamics is provided by an action\footnote{The attentive reader should
notice that in this section the quantity  $F_{\mu\nu}$ represents the
Electromagnetic field.}
\beq
S = \int \sqrt{-\gamma} L(F) d^{4}x
\protect\label{z1}
\eeq
where the Lagrangian $L$ that depends non-linearly on the invariant
$F$ constructed with the field $F_{\mu\nu}$ by the
product\footnote{We consider here, just for simplicity the particular
form of the theory which does not contain the invariant constructed
with the dual of the field.}
\begin{displaymath}
F \equiv F_{\mu\nu} F^{\mu\nu}
\end{displaymath}

The corresponding equation of motion that follows is\footnote{We remind the
reader that although we deal here with Minkowski background metric, we are
using covariant derivatives just in order to exhibit the general covariance of
physics.}

\beq
\left\{L_{F} F^{\mu\nu} \right\}_{;\nu} = \frac{1}{4} J^{\mu}
\label{BI2}
\eeq
where $L_{F}$ represents the functional derivative of the
Lagrangian with respect to the invariant. Maxwell theory is the
case in which this derivative is the constant $- \frac{1}{4}$.

We will limit our analysis here to the theory that represents a non-linear
electrodynamics, which was suggested by Born and developed by Infeld many
years ago. The Lagrangian is given by

\beq
L_{B} = -\frac{1}{4}  \left\{ \sqrt{b^{4} + 2 b^{2} F} - b^{2} \right\},
\protect\label{BI1}
\eeq
in which the constant $b$ has the meaning of the maximum possible
value of the field. There are two important properties of such theory
that interest us, to wit:

\begin{itemize}
 \item{The theory is non-linear.}
 \item{The propagation of the electromagnetic waves can be described
as if the metrical properties of the spacetime were changed by the
presence of the non-linear electromagnetic field.}
\end{itemize}

These two qualities of this type of theory will be explored in the
sequence in order to construct a non-linear spin-two field theory.

Using the general form of expressing the energy-momentum tensor
$T_{\mu\nu}$, presented in a previous section, we obtain from
the Lagrangian eq (\ref{BI1}) the following:

\beq
T_{\mu\nu} = - L \gamma_{\mu\nu} - 4 L_{F} F_{\mu\alpha} {F^{\alpha}}_{\nu}
\label{BI3}
\eeq

We note that this quantity has basically all algebraic properties
and symmetries that appear in the linear Maxwell case. For our purposes here
the interesting property in the non-linear case is the fact that
the interaction of the field with external currents remains the same.
This can be seen most easily by a direct
evaluation of the exchange of field\rq s energy with its sources.
Since we are dealing in the present paper with a most complex situation for the
case of spin-two field, let us spend some time here and show this simple
result in detail in the simpler case of spin-one field, just in order
to get some insight of what should be expected for the spin-two case.

Thus, our task now is to evaluate the ratio of exchange of energy of the field,
that is, the divergence of the symmetric energy-momentum tensor $T_{\mu\nu}$
of the class of non-linear spin-one field,. From the equation of
motion (\ref{BI2}) contracted with $F_{\alpha\mu}$ we obtain

\begin{displaymath}
\left\{L_{F} F_{\alpha\mu} F^{\mu\nu} \right\}_{;\nu}
- L_{F} F_{\alpha\mu;\nu} F^{\mu\nu}   = \frac{1}{4} F_{\alpha\mu} J^{\mu}
\end{displaymath}

Using the expression of the tensor $T_{\mu\nu}$ we re-write this
under the form

\begin{displaymath}
{T^{\alpha\mu}}_{;\mu} + L_{F} F_{,\alpha} + 4 L_{F} F^{\mu\nu}
F_{\alpha\mu;\nu} =  - F_{\alpha\mu} J^{\mu}
\end{displaymath}

and thus finally

\beq
{T^{\alpha\mu}}_{;\mu} = - F_{\alpha\mu} J^{\mu}
\protect\label{BI4}
\eeq

The remarkable fact that follows from this expression is the
well-known result that the balance of forces trough the exchange of
energy of the field and the currents is independent of the form of
the dependence of the Lagrangian on the invariant $F$. This is the
lesson we learn from this simple analysis. Let us pass now to the
gravitational field.

\subsection{A Class of Non-linear Spin-Two Theory}
\label{class}

As we saw in the previous section, when passing from the linear theory of
gravity to the general case the standard
procedure is to add to the energy-momentum tensor of matter, the
corresponding energy tensor for the gravitational field. Proceeding step by
step, the first non-linear term contains $T^{(1)}_{\mu\nu}$, which is the
energy-momentum tensor obtained from the linear part. This procedure is
based on the implicit hypothesis that one should treat the gravitational
energy in the same foot as any other form of energy. In other words, the
gravitational field generated by the gravitational energy is not distinct
from the field generated by any other form of energy. This is a further
extrapolation of the Equivalence Principle, applied to gravitational
energy. At this point we take a path which is different from
the one followed by Feynman, Deser \cite{Feynman}, \cite{Deser}, and others.
Instead of adding to the source of the field the successive
energy-momentum tensors of the gravitational field for each order of
non-linearity, we make the hypothesis that these terms that represent
gravity-to-gravity interaction must be constructed as a functional of
the two invariants $A$ and $B$.
We thus set our action for the free gravitational field to be given by

\beq
S = \int   \sqrt{-\gamma}  L(A, B) d^{4}x
\label{n1}
\eeq

Variation of the potential $\varphi_{\mu\nu}$ yields

\beq
\delta S = \int   \sqrt{-\gamma} \hspace{0.9mm}
{\Theta^{\lambda}}_{\mu\nu;\lambda} \hspace{0.9mm}
\delta \varphi^{\mu\nu} d^{4}x
\label{n2}
\eeq
giving the equation of motion
\beq
{{\Theta}^{\lambda}}_{\mu\nu;\lambda} = 0
\protect\label{n12345}
\eeq
and ${\Theta^{\lambda}}_{\mu\nu}$ is

\beq
\Theta_{\lambda\mu\nu} \equiv 2 L_{A} \left\{ F_{\lambda(\mu\nu)}
- F_{\nu} \gamma_{\lambda\mu} + 2F_{\lambda} \gamma_{\mu\nu} \right\}
- 4 L_{B} \left\{ F_{\mu} \gamma_{\nu\lambda} + F_{\nu}
\gamma_{\mu\lambda} - 2 F_{\lambda} \gamma_{\mu\nu} \right\}
 \label{n4}
\eeq
in which we have used the definitions $L_{A} \equiv \delta L/ \delta A$.
In the special linear case in which $L_{A} = - 1/2$ and
$L_{B} =  3/8$ this expression reduces to the Fierz case:

\beq
{\Theta}^{L}_{\lambda\mu\nu} = - M_{\lambda\mu\nu}.
\label{n5}
\eeq
in which the upperscript $L$ stands for the linear case.
Using the property above (see Section \ref{def}) it follows that
the equation of motion reduces to
\beq
{{M}^{\lambda}}_{(\mu\nu);\lambda} = - 2 {G^{L}}_{\mu\nu} = 0.
\label{n6}
\eeq
and  ${G^{L}}_{\mu\nu}$ is the Fierz linear operator (see eq. (\ref{1})).

Since we would like to impose that our theory should provide the
good weak field limit, that is, Fierz linear equation, we will
restrict our analysis in this paper to those Lagrangians whose
dependence on the invariants obeys the relationship:

\beq
L_{B} = - \frac{3}{4} L_{A}.
\protect\label{x1}
\eeq

Under this hypothesis the equation of motion for the free
gravitational field within our scheme, equation (\ref{n12345}) takes the form

\beq
\left\{ L_{A} {M^{\lambda}}_{(\mu\nu)} \right\}_{;\lambda} = 0.
\protect\label{n7}
\eeq

Using the properties of ${M^{\lambda}}_{(\mu\nu)}$ we can re-write
this expression in a more convenient form:

\beq
{G^{L}}_{\mu\nu} = \frac{1}{2} {L_{A}}^{-1} \left\{ L_{A;\lambda}
{M^{\lambda}}_{(\mu\nu)}\right\}
\protect\label{n8}
\eeq

Note that this is an exact equation, that is, it does not contain any sort of
approximation term. We have just isolated the
linear Fierz operator on the left-hand side
and set all non-linearity terms to the right-hand side. Besides, under this
form one can see directly that the
source of the non-linearity, contrary to the case of GR, is not
expressed in terms of the energy-momentum tensor of the gravitational field.
We will make this
point clearer when we treat the case of interaction with matter.

\subsection{The Gravitational Energy-Momentum Tensor}
\label{energy}

The standard definition of the energy-momentum tensor for any field
is provided by the variation of the Lagrangian with respect to the
underlying metric, through the expression

\beq
{T^{g}}_{\mu\nu} = \frac{2}{\sqrt {-\gamma}} \frac{\delta L
\sqrt{-\gamma}}{\delta \gamma^{\mu\nu}}
\label{n10}
\eeq

For the Lagrangians that we examine here (which obey the condition
set up trough equation (\ref{x1})) we have

\beq
{T^{g}}_{\mu\nu} = - L \gamma_{\mu\nu} + 2 L_{A} \left\{
\frac{\delta A}{\gamma^{\mu\nu}} -
\frac{3}{4}\frac{\delta B}{\gamma^{\mu\nu}} \right\}
\label{n11}
\eeq

After a rather long although direct calculation we obtain the form:

\beq
{T^{g}}_{\mu\nu} = - L \gamma_{\mu\nu} + L_{A} \left\{ 4
F_{\mu\alpha\beta}  {F_{\nu}}^{\alpha\beta} + 2
F_{\alpha\beta\mu}  {F^{\alpha\beta}}_{\nu} - 3 F^{\alpha}
F_{\alpha(\mu\nu)} - \frac{5}{2} F_{\mu} F_{\nu} + F^{\epsilon}
F_{\epsilon} \gamma_{\mu\nu} \right\}
\label{n12}
\eeq
in which the symbol $g$ stands for {\bf gravity}.

This is the form of the gravitational energy-momentum tensor of the
gravitational field, obtained from the Lagrangian $L$.

Take the trace of the above form to arrive at

\beq
T^{g} = - 4 L + 6 L_{A} \left\{ A - \frac{3}{4} B \right\}
\label{n13}
\eeq

It is a direct exercise, left to the reader, to show that in the
particular case of the linear action, (equation (\ref{s1})), this
expression reduces to Gupta energy-momentum tensor.

The tensor ${T^{g}}_{\mu\nu}$ differs from the one obtained from
Noether\rq s theorem by a total divergence. For future references it
is useful to write the Noether energy-momentum tensor:

\beq
N_{\mu\nu} = - L \gamma_{\mu\nu} - 2 \varphi_{\alpha\beta;\nu}\hspace{0.9mm}
 L_{A} \left\{ F_{\mu(\alpha\beta)}  + \frac{1}{2} F^{(\alpha}
\gamma^{\beta)\mu} - F^{\mu}  \gamma^{\alpha\beta} \right\}
\label{n121}
\eeq
or, equivalently

\beq
N_{\mu\nu} = - L \gamma_{\mu\nu} -  \varphi_{\alpha\beta;\nu}\hspace{0.9mm}
 L_{A} M_{\mu\alpha\beta}
\label{n1212}
\eeq

Under this form the examination of the balance of energy between the
gravitational field and its sources assumes a very simple expression

\beq
{{N^{\mu}}}_{\nu;\mu} = 2 T_{\alpha\beta}\hspace{0.9mm}
 {{\varphi}^{\alpha\beta}}_{;\nu}
\label{122}
\eeq
This can be shown either using an analogous trick as in the spin-one
case as described above or just by direct calculation. Let us
remark that, like in the previous case of spin-one, the balance of energy
between the gravitational field and its sources is independent of the form of
the Lagrangian one takes to represent the gravitational field.

In the next section we will consider a specific simple case of
gravitational theory by a choice of the Lagrangian in this scheme.

\section{A SUGGESTIVE MODEL TO GRAVITATION}
\protect\label{model}

We have dealt in the precedent sections with the general scenario for
our construction of a theory of gravity.
The aim of this section is to produce a specific characterization of
the gravitational equations of motion by searching a Lagrangian
that satisfies the requirements
set up in the precedent sections. This means that we will undertake now the
task to produce a specific example of a field theory for the gravitational
field that fulfills both conditions:

\begin{itemize}
 \item{obeys the requirements set up in the precedents section
(including that it satisfies the equation \ref{n7}).}
 \item{agrees with the observed tests of the gravitational field.}
\end{itemize}

Just to simplify our explanation
of the properties of our proposal we will take as a model of our
scenario the non-linear theory of electromagnetic field proposed
by Born and Infeld that we presented above. We do
not intend to solve completely our task of
searching a field theory for the gravitational
field with such a naive model,
but only to present our general scheme to deal with this question.

Taking this theory as an example, let us assume as a toy model for
the self-interacting gravitational field the Lagrangian

\beq
L^{g} = \frac{1}{\kappa} \left\{ \sqrt{b^{4} +  b^{2} (-A + \frac{3}{4} B) }
 - b^{2} \right\}
\label{BI12}
\eeq

The quantity $b$ has the dimension of $(lenght)^{-1}$. At this level of
the theory it is a free parameter that can be choose either by some
speculative consideration (e.g.by setting it equal to Planck\rq s
lenght, for instance) or by future observational requirements. By the
time being we will left it free. Let us remark that, for the class of
Lagrangians (eq. (\ref{BI12})), this quantity does not have the same meaning
as in the original Born proposal of the maximum value possible for the field.

In the present presentation we will concentrate our interest on the
consequences of such theory in the propagation of the gravitational
waves. We leave the exam of the consequences of this model
to a subsequent paper\footnote{See Conclusions for other remarks
related to the properties of this theory.}.

\subsection{GRAVITATIONAL WAVES}
\label{waves}

The main purpose of this section is to examine the
behavior of the gravitational waves in this theory. In order to do
this we will analyse the evolution of discontinuities of the equation of
motion through  a characteristic surface $\Sigma$. This analysis give us the
velocity of the gravitational wave that will be the key point to
distinguish this kind of theory from General Relativity. For pedagogical
reasons we start by examining the non-linear spin-one case.

\subsubsection{Propagation of the Discontinuities: The Non-Linear Spin-1 Case}

Let $\Sigma$ be the surface of discontinuity. We set (using Haddamard's
condition)

\beq
[F_{\mu\nu}]_{\Sigma} = 0,
\label{gw1}\eeq
and

\beq
[F_{\mu\nu,\lambda}]_{\Sigma} = f_{\mu\nu}{\bf k}_{\lambda}.
\label{gw2}\eeq
in which the symbol $[ J ]_{\Sigma}$ represents the discontinuity
of the function $J$ through the surface $\Sigma$.

Applying this into the equation of motion (\ref{BI2}) we obtain

\beq
L_{F}f^{\mu\nu}{\bf k}_{\nu} + 2 L_{FF}\eta F^{\mu\nu}\nk =
0,
\label{gw3}\eeq
where $\eta$ is

$$ F^{\alpha\beta}f_{\alpha\beta} \equiv \eta. $$

After some algebraic manipulations the equation of propagation of the
disturbances is provided by

\beq
\left\{\gamma_{\mu\nu}( \frac{L_{F}^{2}}{L_{FF}} + L ) +
T_{\mu\nu} \right\} k^{\mu} k^{\nu} = 0
\label{gw4}
\eeq

In the case of the Born theory this expression reduces to

 \beq
\left\{\gamma_{\mu\nu} + \frac{4}{b^{2}}T_{\mu\nu} \right\}  k^{\mu} k^{\nu}
= 0
\label{gw5}
\eeq

In this particular non-linear non-polynomial theory we see that
the disturbances propagate in the modified geometry,
changing the background geometry $\gamma_{\mu\nu}$ into an effective
one $g_{\mu\nu}$, which depends on the energy distribution of the field.
We shall see later that the same structure occurs for spin-2.

\subsubsection{Propagation of the Discontinuities: The Non-Linear Spin-2 Case}
\label{spin-2}

In an analogous way we set for the spin-2 field the discontinuity
conditions

\beq
[F_{\mu\nu\alpha}]_{\Sigma} = 0
\label{gw6}\eeq
and

\beq
[F_{\mu\nu\alpha;\lambda}]_{\Sigma} = f_{\mu\nu\alpha}{\bf k}_{\lambda}
\label{gw7}\eeq

that imply

\beq
[M_{\mu\nu\alpha;\lambda}]_{\Sigma} = m_{\mu\nu\alpha}{\bf k}_{\lambda}
\label{gw71}\eeq
which is the compact form that appears in the equation of motion  (\ref{n7}).
Taking the discontinuity of this equation we obtain

\beq
m_{\mu(\alpha\beta)} \mk + \frac{4}{b^2}L_{A}^{2} (\eta -
\frac{3}4 \Lambda) M_{\mu(\alpha\beta)} \mk = 0
\label{gw8}
\eeq
in which the quantities $\eta$ and $\Lambda$ are defined by

$$ \eta \equiv F_{\alpha\beta\mu} f^{\alpha\beta\mu} $$

$$ \Lambda \equiv M_{\mu} m_{\mu} $$

After some algebraic manipulations (see the Appendix for some
details) the equation of propagation of the disturbances is written as

\beq
\left\{\gamma_{\mu\nu}(\frac{b^{2}}{2L_{A}} + L ) +
{T^{g}}_{\mu\nu} \right\}  k^{\mu} k^{\nu} = 0
\label{gw9}
\eeq

In the case of the Born-like Lagrangian (see equation (\ref{BI12}) )
this expression reduces to

 \beq
\left\{\gamma_{\mu\nu} - \frac{1}{b^{2}} {T^{g}}_{\mu\nu} \right\}
 k^{\mu} k^{\nu} = 0
\label{gw10}
\eeq
in which  ${T^{g}}_{\mu\nu}$ is the energy-momentum tensor of the
gravitational field defined previously.

We note that in our theory we face a very similar behavior as in the
previous spin-1 case. Indeed, the disturbances propagate
in a modified geometry,
changing the background geometry $\gamma_{\mu\nu}$, into an effective
one $\tilde{g}_{\mu\nu}$ which depends on the energy distribution of the field
$F_{\alpha\beta\mu}$. This fact shows that such a property stems from
the structural form of the Lagrangian.

{}From the above calculations we conclude that, differently from General
Relativity, in the present theory the caracteristic surfaces of the
gravitational waves propagate on the null cone of
an effective geometry distinct of that observed by {\bf all} other forms of
energy and matter. This result gives a possibility to
choose between these two theories just by observations of the gravitational
waves. This is a challenge that is expected to be solved in the near future.

\section{THE GRAVITY-MATTER INTERACTION}
\label{matter}

As we pointed out in Section (\ref{introduction}) there are strong
evidences that matter couples universally with gravitation in such a
way that its net effect can be described as if matter is imbedded
in a Riemannian geometry produced by the gravitational field. Many
authors \cite{Deser} \cite{Gris} have shown that this geometry can be
written in terms of an unobservable background geometry  $\gamma^{\mu\nu}$
and the gravitational field through the potential $\varphi^{\mu\nu}$
as\footnote{For some practical simplifications some authors do not deal
with such a definition, but instead with pseudo-tensors obtained by
multiplication of these tensors by the square-root of the corresponding
determinants. See, for instance, \cite{Gris}.}

\beq
g^{\mu\nu} \equiv \gamma^{\mu\nu} + \varphi^{\mu\nu}
\protect\label{GM1}
\eeq

This means that in order to know how matter couples to gravity
one must just make the substitution of the background geometry
$\gamma^{\mu\nu}$ by an effective geometry $g^{\mu\nu}$
and change accordingly the derivatives by the covariant derivatives
constructed with $g^{\mu\nu}$. Since such procedure is already a standard
one and there are very good reviews on this in the scientific literature we
will not enter here in more details on this field theoretical
equivalence of description of General Relativity.
The reader that is not familiar with this should consult the review
done in \cite {Gris}.

Let us just take here as a very simple example, the case of a
scalar field $\Psi$. The free field equation of motion is provided by the
action

\beq
S = \int \sqrt{-\gamma} \Psi_{,\mu} \Psi_{,\nu} \gamma^{\mu\nu} d^{4}x
\label{e1}
\eeq

In order to couple the scalar matter field with gravity we make the
substitution of the metric $\gamma^{\mu\nu}$ into the effective one defined
as above. The action takes then the form

\beq
S = \int \sqrt{-g} \Psi_{,\mu} \Psi_{,\nu} (\gamma^{\mu\nu} +
\varphi^{\mu\nu}) d^{4}x
\label{e3}
\eeq
that is

\beq
S = \int \sqrt{-g} \Psi_{,\mu} \Psi_{,\nu} g^{\mu\nu} d^{4}x
\label{e4}
\eeq

It seems worth to remark that in this expression we have set the
determinant of the inverse

$$ g \equiv det\hspace{0.7mm} g_{\mu\nu} = {det (\gamma^{\mu\nu} +
\varphi^{\mu\nu})}^{-1}. $$
Note that the quantity  $g_{\mu\nu}$ is the inverse of  $g^{\mu\nu}$
and constitutes an infinite series in terms of the tensors
$\gamma^{\mu\nu}$ and $\varphi^{\mu\nu}$.
Then the equation of motion of the scalar field becomes

\beq
\Box \Psi \equiv \frac{1}{\sqrt{-g}} \left\{ \sqrt{-g} \Psi_{,\mu}
g^{\mu\nu} \right\}_{,\nu}  =  0.
\label{4321}
\eeq

We can thus obtain the modification of the equation of motion of the
gravitational field in presence of matter. In the above hypothesis
(borrowed from General Relativity) that the matter feels the
gravitational field only by the combination
$g^{\mu\nu} \equiv \gamma^{\mu\nu} + \varphi^{\mu\nu}$ it follows
that

$$ \frac{\delta L}{\delta \gamma^{\mu\nu}} =
\frac{\delta L}{\delta g^{\mu\nu}} $$

Thus in the theory developed here the general equation of motion of
the gravitational field containing source terms takes the form

\beq
\left\{ L_{A} {M^{\lambda}}_{(\mu\nu)} \right\}_{;\lambda} = - T_{\mu\nu},
\protect\label{n171}
\eeq

or the equivalent one

\beq
{G^{L}}_{\mu\nu} = \frac{1}{2} {L_{A}}^{-1} \left\{ L_{A;\lambda}
{M^{\lambda}}_{(\mu\nu)} + T_{\mu\nu} \right\}
\protect\label{n181}
\eeq

in which $T_{\mu\nu}$ represents the matter energy-momentum tensor of
matter evaluated by the equation (\ref{n10}).

Finally, we would like to emphasize that, as we have pointed out before,
we can thus conclude that, in what concerns
the matter to gravity interaction the theory proposed here is
indistinguishable of the General Relativity.

\section{CONCLUSION}
\label{conclusion}
Einstein\rq s theory of General Relativity is one of the most beautiful,
comprehensive and deeply important achievements of classical field theory.
Not only it contains and exhibits such a simplicity and internal coherence
but, more than this, it provides a sound step towards the understanding of
gravitational processes which has had no rival since the early times of
Newton hypothesis concerning the existence of the universal gravitational
atraction. Thus, any theory that dares to propose even a small
modification on its scenario faces not only the challenge to present
a sound new argument -but besides, it has to deal with the enormous difficulty
of intending the substitution of a successful paradigm.

Thanks to the work of many physicists as Feynmann, Deser and others,
we have learned that Einstein\rq s  geometric vision of General Relativity
admits an equivalent field theoretical presentation.  In such a formulation,
the phenomenon of gravity becomes more similar to the rest of Physics and,
besides (and by far the more important) we obtain a deeper comprehension on
the mechanism of self-interaction of the gravitational field. It is precisely
the analysis of this mechanism that we think is worth to re-examine.
We can thus realize that there are two reasons for this:

\begin{itemize}
 \item{The theoretical challenge of providing another coherent manner to
describe gravitational self-interaction process.}
 \item{The perspective of test alternative theories by the
observation, in the near future, of gravitational waves.}
\end{itemize}

In the present paper we describe a new program of analysis of gravitational
interaction that has a very similar geometrical interpretation as
General Relativity as far as matter to gravity
processes are concerned. However, it becomes distinct from GR in the
description of gravity to gravity interaction. We have exhibited a specific
example of a theory in order to show how these ideas can be implemented.
We have evaluated the propagation of gravitational waves within such a theory
and shown that the velocity of the waves does not coincide with the GR
prediction. Once there is a large expectative that these waves will
be detected in the next few years, we could test this aspect of
these theories soon.

Finally, we should mention that we have found two particular
solutions of the theory we present here: the spherically symmetric
static case and a cosmology. We can anticipate that both solutions
are in good agreement to actual observations. Both will be published soon.

\section{APPENDIX: SOME USEFUL FORMULAE}

\subsection{The Traceless Part of the Gravitational Field}

The traceless tensor $C_{\alpha\mu\nu}$ is defined by

$$  C_{\alpha\mu\nu} \equiv F_{\alpha\mu\nu} - \frac{1}{3}
F_{\alpha} \gamma_{\mu\nu} + \frac{1}{3} F_{\mu} \gamma_{\alpha\nu} $$

The invariant constructed with this quantity $C$ is given by

$$ C \equiv  C_{\alpha\mu\nu} C^{\alpha\mu\nu} $$

It then follows that between this quantity and the invariants dealt with in
the paper ($A$ and $B$)there is the following relation

$$ C = A - \frac{2}{3} B. $$

We could then use the irreducible parts of the gravitational field to describe
its dynamics. In this case the generic form of the Lagrangian should
be

$$ L [ B, C ]. $$.

\subsection{Variational Formulae}

The evaluation of the energy-momentum of the gravitational field needs
some technicalities concerning the dependence of $F^{\alpha\mu\nu}$
on the background metric $\gamma_{\mu\nu}$. There are some expressions that
simplify this manipulation. Let us present here some of these.

We have

$$ \delta A = \left\{ 2 F_{\mu\alpha\beta}  {F_{\nu}}^{\alpha\beta} +
F_{\alpha\beta\mu}  {F^{\alpha\beta}}_{\nu} + \frac{3}{2}  F^{\alpha}
F_{\alpha(\mu\nu)} + F_{\mu} F_{\nu} - F^{\epsilon} F_{\epsilon}  \right\}
\delta \gamma^{\mu\nu} $$

To show the validity of this expression we must use the relations

$$ F^{\alpha\beta\mu}  \delta F_{\alpha\beta\mu} = \frac{1}{2}
F^{\lambda} \gamma^{\beta\mu} \delta F_{\lambda\beta\mu} $$
and

$$ F^{\lambda} \gamma^{\beta\mu} \delta F_{\lambda\beta\mu} =
\left\{ \frac{3}{2}  F^{\alpha} F_{\alpha(\mu\nu)} +
F_{\mu} F_{\nu} - F^{\epsilon}
F_{\epsilon}  \gamma_{\mu\nu} \right\} \delta \gamma^{\mu\nu}  $$

For the invariant $B$ we obtain

$$ \delta B =  \left\{ 4 F^{\alpha} F_{\alpha(\mu\nu)} + 3 F_{\mu} F_{\nu} -
2 F^{\epsilon} F_{\epsilon}  \gamma_{\mu\nu} \right\}
\delta \gamma^{\mu\nu}  $$

\subsection{Algebraic Formuli}

We list some useful relations that are worth obtaining in order to deal with
the quantities presented in the text.

A simple inspection on the equations of motion of the dynamics of the
present theory shows that it depends only on the symmetric part of the
two last indices of the gravitational field. This is a consequence of the
symmetry of the potential $\varphi_{\mu\nu}$. Note however that even the
invariants of the theory have already this property. Indeed, we can write

$$ A \equiv F_{\alpha\mu\nu} F^{\alpha\mu\nu}
 = \frac{1}{3} F_{\alpha(\mu\nu)} F^{\alpha(\mu\nu)}.  $$

Another consequence of this symmetry is explicitated through the
expression

$$  F_{\alpha(\mu\nu)} + F_{\mu(\alpha\nu)} + F_{\nu(\alpha\mu)} = 0. $$

Consequently, the same formula is valid for the associated quantity
$M_{\alpha\mu\nu}$, that is,

$$  M_{\alpha(\mu\nu)} + M_{\mu(\alpha\nu)} + M_{\nu(\alpha\mu)} = 0. $$

These expressions allow us to re-write the gravitational
energy-momentum tensor ${T^{g}}_{\mu\nu}$ under the equivalent form

$$ {T^{g}}_{\mu\nu} = - L \gamma_{\mu\nu} + \frac{1}{3} L_{A} \left\{ 2
M_{\mu(\alpha\beta)}  {M_{\nu}}^{(\alpha\beta)} + 4
M_{\alpha(\beta\mu)}  {M^{\alpha}}_{(\beta\nu)} + 3 M^{\alpha}
M_{\alpha(\mu\nu)} - \frac{3}{2} M_{\mu} M_{\nu} \right\}  $$

There is a cyclic identity\footnote{See, for instance \cite{Duque}}
concerning the first derivative of the field which is given by

$$ {{F_{\lambda\alpha}}^{\nu}}_{;\beta} +
{{F_{\alpha\beta}}^{\nu}}_{;\lambda} +
{{F_{\beta\lambda}}^{\nu}}_{;\alpha} = \frac{1}{2} \left\{
\delta^{\nu}_{\alpha} A_{[\beta\lambda]} + \delta^{\nu}_{\beta}
A_{[\lambda\alpha]} + \delta^{\nu}_{\lambda} A_{[\alpha\beta]}
\right\} $$

in which we have defined

$$ A_{\mu\nu} \equiv - {F^{\alpha}}_{\mu\nu;\alpha} - F_{\mu;\nu} $$

In terms of the quantity $M^{\alpha\beta\nu}$ this expression then takes
the form

$$ {{M_{\lambda\alpha}}^{\nu}}_{;\beta} +
{{M_{\alpha\beta}}^{\nu}}_{;\lambda} +
{{M_{\beta\lambda}}^{\nu}}_{;\alpha} = $$
$$ \frac{1}{2}
\delta^{\nu}_{\alpha} \left\{ {{M_{\beta\lambda}}^{\epsilon}}{;\epsilon} -
  \frac{1}{2} M_{[\beta\lambda]} \right\} +
\frac{1}{2} \delta^{\nu}_{\beta} \left\{
{{M_{\lambda\alpha}}^{\epsilon}}{;\epsilon} - \frac{1}{2} M_{[\lambda\alpha]}
\right\} + \frac{1}{2} \delta^{\nu}_{\lambda} \left\{
{{M_{\alpha\beta}}^{\epsilon}}{;\epsilon} - \frac{1}{2} M_{[\alpha\beta]}
 \right\} $$

\end{document}